\newcommand{\CLASSINPUTinnersidemargin}{18mm}
\newcommand{\CLASSINPUToutersidemargin}{12mm}
\newcommand{\CLASSINPUTtoptextmargin}{20mm}
\newcommand{\CLASSINPUTbottomtextmargin}{25mm}
\documentclass[10pt,conference,twoside,a4paper]{IEEEtran}
\usepackage{tabularx}

\usepackage[utf8]{inputenc}

\usepackage{times}

\renewcommand{\abstractname}{Abstract}    

\usepackage{graphicx}
\usepackage{subfigure}
\usepackage{svg}
\DeclareGraphicsExtensions{.png,.eps,.ps,.pdf,.svg}

\usepackage{fancyhdr}
\usepackage{kantlipsum}

\usepackage{url}

\usepackage[left=2cm,top=2.5cm,right=2cm,bottom=2.5cm]{geometry}

\usepackage[]{color}

\usepackage{eso-pic}

\begin{document}
\AddToShipoutPictureBG*{%
  \AtPageUpperLeft{%
    \setlength\unitlength{1in}%
 	\hspace{2cm}
 	 	\makebox(0,-2)[l]{
			\begin{tabular}{l r} 
			\multicolumn{1}{p{12cm}}{\vspace{-0.3cm}\includegraphics[width=1.1\columnwidth]{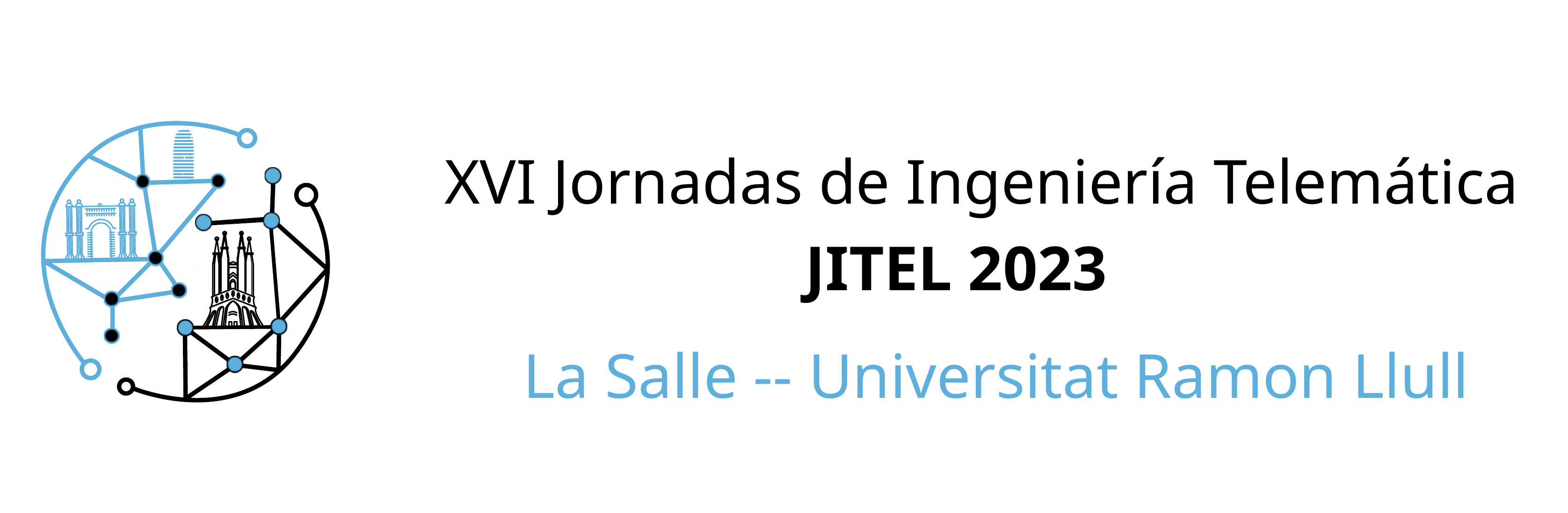}} & \multicolumn{1}{p{4cm}}{\raggedleft\small\usefont{T1}{phv}{m}{it} Actas de las XVI Jornadas de Ingeniería Telemática\\ (JITEL 2023),\\ Barcelona (España), \\8-10 de noviembre de 2023. \vspace{0.2cm} \\ISBN: 978-84-09-58148-1} \tabularnewline 
			\end{tabular}
 }
}}

\AddToShipoutPictureBG*{%
  \AtPageLowerLeft{%
    \setlength\unitlength{1in}%
    \hspace*{\dimexpr0.5\paperwidth\relax}
    \makebox(0,1.3)[c]{\footnotesize\usefont{T1}{phv}{m}{} This work is licensed under a \underline{\textcolor{blue}{Creative Commons 4.0 International License}} (CC BY-NC-ND 4.0)}%

}}

\title{\vspace{2cm}SareQuant: Towards a quantum-based communication network}

\author{\IEEEauthorblockA{Ane Sanz$^{1,2}$, David Franco$^1$, Asier Atutxa$^1$, Jasone Astorga$^{1,2}$, Eduardo Jacob$^{1,2}$}
\IEEEauthorblockA{$^1$Department of Communications Engineering, University of the Basque Country (UPV/EHU). 48013 Bilbao, Spain.\\
$^2$EHU Quantum Center, University of the Basque Country (UPV/EHU). 48940 Leioa, Spain.\\
\{ane.sanz, david.franco, asier.atutxa, jasone.astorga, eduardo.jacob\}@ehu.eus}
}

\maketitle

\begin{abstract}
\textbf{This paper presents the SareQuant project, which aims to evolve the Basque NREN (National Research and Education Networks) into a quantum-based communication infrastructure. SareQuant focuses on the network design and on the integration of quantum technologies into real-world scenarios and applications. Therefore, this paper provides insights into the opportunities and challenges regarding the integration of quantum technologies, thus paving the way for a secure and advanced Quantum Internet.}
\end{abstract}


\section{\uppercase{Introduction}}
In today's rapidly evolving digital era, National Research and Education Networks (NRENs) play a crucial role in supporting different research and education activities. NRENs encompass high-speed networks that interconnect multiple research and education institutions, such as universities and research centers, to provide high-speed connectivity and access to advanced resources. NRENs also act as the fundamental framework for knowledge sharing, enabling seamless collaboration, resource sharing, and use of advanced technologies among researchers and students. As an example, the Global P4 Lab (GP4L) is a high-performance communication network for research and education that leverages resources of the European NRENs' for their interconnection.
Considering the continuous and rapid evolution of technology, there is a strong need for NRENs to enhance their capabilities by adopting and integrating emerging technologies. 

Among the emerging technologies, Quantum Technologies stand out as one of the most promising for next-generation services and applications, including Quantum Computing and Quantum Communications. Such technologies, although still in early stages of development and lacking full maturity, are expected to have a major impact and challenge conventional systems. Quantum Computing, for instance, promises to solve highly complex operations that are beyond the capabilities of classical computing. Quantum communication, on the other hand, offers information-theoretic security by leveraging the principles of quantum mechanics. In addition, the integration of such technologies, as well as the interconnection of quantum devices, implies the deployment of quantum networks that must coexist with classical ones, forming hybrid classical-quantum networks that support new capabilities \cite{van2014quantum}. It is envisaged that these networks, which will be deployed gradually based on the availability and maturity of the technology, will culminate in a network commonly referred to as the Quantum Internet (QInternet).

In this context, it is important to note that, due to the intrinsic characteristics of quantum technologies, the architecture and operation of the QInternet vastly differs from the classical one, which means that the transition to a quantum-enabled network is not immediate and needs to be studied in more detail. 

Therefore, this paper introduces the current status of the SareQuant project, which aims at analysing the requirements and implications for the evolution of I2Basque, the NREN of the Basque Country, towards a quantum-based network architecture. On the other hand, SareQuant also seeks the integration of quantum technologies into real-world scenarios, employing key technologies identified for the initial stages of the Quantum Internet.


\section{\uppercase{Quantum Internet}}
This section describes the concept of QInternet, exploring the identified opportunities and challenges. It also provides a description of the Quantum Key Distribution (QKD) technology, identified as a key enabler for the early stages of development of the QInternet.

\subsection{Concept, opportunities and challenges}
The QInternet is expected to enhance classical Internet by enabling quantum communications between any two points in the world, which will in turn enable new services and applications beyond the scope of classical networks, such as the realisation of complex optimisation problems. According to the Internet Research Task Force (IRTF), which has published several RFC and drafts regarding the QInternet \cite{rfc9340}, \cite{irtf-draft-use-cases}, the deployment of the QInternet involves the definition of a new quantum network stack that accounts for fundamental principles of quantum mechanics such as superposition, entanglement or measurement. 

\begin{figure}[t]
\centerline{
\includegraphics[width=\columnwidth]{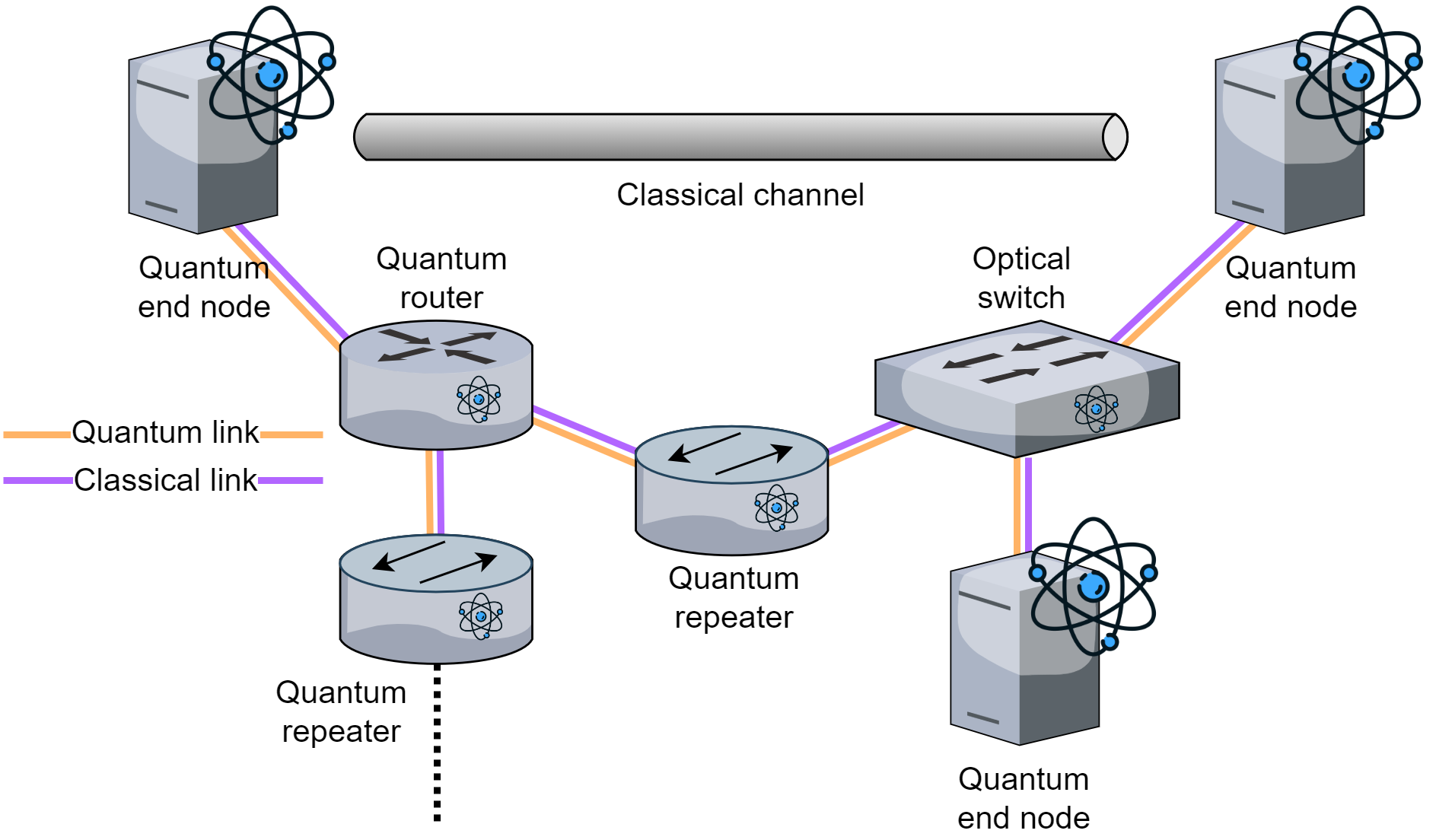}
}
\caption{Quantum Internet generic architecture.}
\label{fig:QI}
\end{figure}

Additionally, the RFC 9340 defines the architectural principles of the QInternet, identifying, among others, the elements that comprise the network. According to this RFC, the main elements that should make up the network are quantum routers, quantum repeaters, quantum end-nodes, and passive elements such as optical switches. In addition, most quantum devices require of both quantum and classical links to properly perform their processes and tasks. Fig. \ref{fig:QI} shows a generic quantum network architecture where all the above elements are represented. This scheme represents a scenario where two applications running on two end nodes with quantum capabilities need to communicate with each other, which requires the establishment of an end-to-end connection through different quantum devices.

However, it is important to outline that the deployment of quantum networks is conditioned by the availability of the involved technologies and devices. In fact, quantum repeaters, which are considered essential for establishing long-range links, remain unavailable at the moment. This means that it is not possible to completely stick to the presented generic architecture, and network configurations must be adapted to the maturity and availability of the technology at any given time. For instance, the interconnection of different quantum devices presently relies on a combination of both optical switches and trusted relays that enable connections over medium-large distances. The eventual availability of quantum repeaters will presumably increase this capability, further enhancing the interconnection of quantum devices. 

Therefore, there exists a need for a progressive evolution of conventional networks towards a quantum-based infrastructure, in line with technological advances. In this context, studies such as \cite{wehner2018QI} concur with the notion of developing the QInternet in different stages, with the first stages focused on the implementation of QKD systems. 

\subsection{Quantum Key Distribution as an early stage}
QKD technology enables the generation of a symmetric key between two endpoints in a information-theoretic secure manner, ensuring the utmost privacy and confidentiality of such key. The underlying principle of QKD, which combines the use of both quantum and classical channels, is that any attempt to observe the transmitted photons in the quantum channel disturbs the transmission in a way that induces detectable errors at both communication ends. Thus, considering the ability of establishing and distributing secure keys, QKD emerges as one of the most promising quantum communication technologies for achieving ultra-secure communications \cite{Survey2018QKD}.

There are two main implementation options available for QKD systems, based on the information encoding method: Discrete-Variable (DV-QKD) and Continuous Variable (CV-QKD). In DV-QKD systems, discrete quantum states such as polarization or phase are used for encoding information. Conversely, CV-QKD systems employ continuous variables of quantum states, such as the quadrature components of the electromagnetic field for information encoding. Additionally, QKD systems can employ two main approaches for the transmission and measurement of quantum signals, namely prepare-and-measure and entanglement-based. In the prepare-and-measure approach, the transmitter is responsible for preparing and transmitting the quantum states, while the receiver undertakes the task for measuring them. On the other hand, entanglement-based approaches entail the use of an external source that generates and distributes entangled photons between both communication ends for subsequent measurement. 

Therefore, QKD is currently one of the quantum technologies with the highest level of maturity, supported by the availability of commercial equipment from multiple international manufacturers. This accessibility of QKD devices greatly facilitates research in the area and the implementation of the technology. Consequently, the advanced stage of development and availability of QKD, as well as its inherent advantages in terms of unconditional security, position it as the most suitable technology for early-stage deployment in the QInternet.


\section{\uppercase{Quantum Initiatives in Spain}}
The quantum technologies sector has gained significant interest in the last years from both companies and public institutions within the international community. This growing attention can be reflected in the multitude of initiatives that have been launched with the aim of promoting the advancement of such technologies.

In the Spanish context, the \textit{Plan Complementario de Comunicación Cuántica} launched by the Ministry of Science and Innovation, has earmarked 54 million EUR to fund research projects that foster the development of quantum digital technologies to enforce the cybersecurity in Spain. This initiative aligns with other European initiatives with similar objectives, such as the EuroQCI or the OpenQKD. Specifically, within the scope of these projects, some work has already been done over local NRENs, as the MadQCI infrastructure \cite{MadQCI}, a quantum-based metropolitan network, which is undergoing its development.

At the regional level, the Basque Country has also launched the IKUR strategy that prioritises four specific fields, including the Quantum Technologies. This strategy aims at attracting and generating quantum-related talent, fostering the development of novel infrastructures and positioning the Basque Country as an international leader. 

Similarly, the University of the Basque Country (UPV/EHU) has recently created the EHU Quantum Center. The objective of this center, which is considered itself an actor of the IKUR strategy, is to coordinate members, groups, and activities, to contribute to scientific excellence, and to participate in public and private quantum initiatives.

Therefore, all these initiatives demonstrate the growing interest of local, national, and international institutions in fostering the development of quantum technologies. In addition, the SareQuant project presented in this work, is realised in the context of the aforementioned initiatives, specifically as a part of the IKUR initiative, contributing to the collective efforts aimed at advancing quantum technologies and their applications.

\section{\uppercase{SareQuant: towards a quantum-based network}}
SareQuant is a project aimed at proposing and designing an infrastructure compatible with the existing I2Basque, the basque NREN, to allow experimentation with quantum technologies, taking a first step in the evolution towards a QInternet. This infrastructure would extend the network's current functionalities, transforming it into a practical testbed for using and evaluating quantum technologies and supporting experimentation with already accessible quantum technologies in the initial phases. Therefore, SareQuant encompasses two main lines of work. The first one is focused on the progressive evolution of the actual infrastructure design towards a QInternet, while the second concentrates on the integration of quantum technologies, specifically QKD, into real-world scenarios.

\subsection{Current state of I2Basque}
\begin{figure}[t]
\centerline{
\includegraphics[width=\columnwidth]{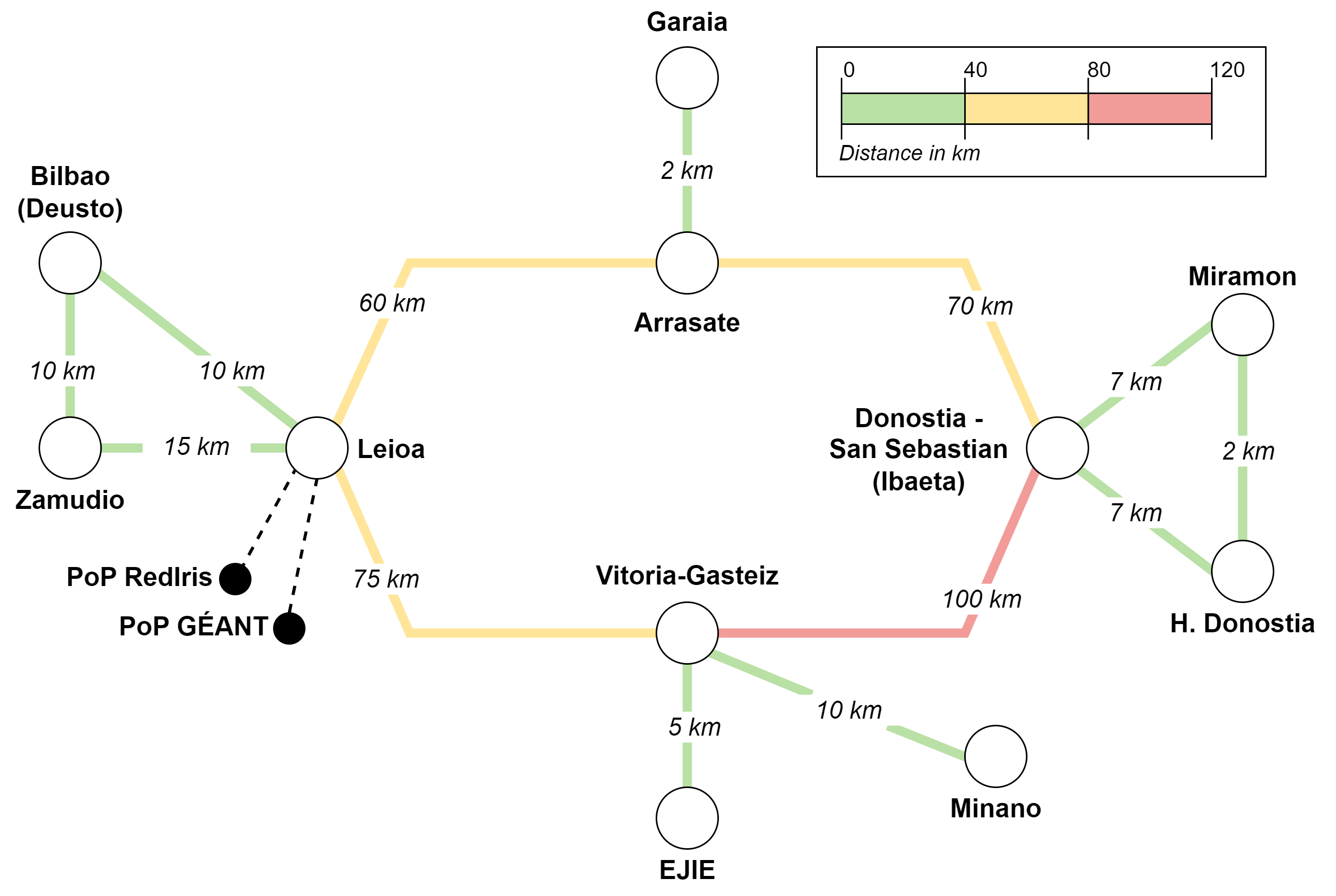}
}
\caption{I2Basque network nodes and links}
\label{fig:I2Basque}
\end{figure}

I2Basque is responsible for providing advanced network infrastructure and services to the research and education communities of the region. This network, as depicted in Fig. \ref{fig:I2Basque}, consists of a central ring with four main Points of Presence (PoPs) in Donostia-San Sebastián, Arrasate, Leioa and Vitoria-Gasteiz, connected through 10 Gbps lambda links. This interconnection leverages Dense Wavelength Division Multiplexing (DWDM) technology to allow simultaneous transmission of multiple wavelengths over a single optical fiber. It also includes two metropolitan rings with dark fiber in Bizkaia between Bilbao-Leioa-Zamudio, and in Donostia between Ibaeta-Miramon-H.Donostia, as well as different dark fiber links in Gasteiz and Arrasate. The I2Basque network topology comprises several nodes connected by links of different distances, all of them within an acceptable range in terms of feasibility with current quantum technology, as shown in Fig. \ref{fig:I2Basque}. This feature, which holds significant importance and often becomes critical in the context of quantum communications, positions this network as an ideal platform for initiating the transition towards a quantum-based network. 

Furthermore, this infrastructure also has PoPs to the networks of RedIris, the Spanish NREN, and GÉANT, the collaboration of European NRENs. This PoPs offer several benefits to the I2Basque network, including improved and direct collaboration opportunities with other international research and education institutions present in GÉANT, thereby expanding the potential for experimentation.

\begin{table*}[t]
\centering
\caption{Key factors for the integration of QKD into real-world applications}
\label{tab:tabla}
\resizebox{\textwidth}{!}{%
\begin{tabular}{
| m{0.13\textwidth}
| m{0.35\textwidth}
| m{0.39\textwidth} | }
\hline
\multicolumn{1}{|c|}{\textbf{Factor}} & 
\multicolumn{1}{c|}{\textbf{Description}} & 
\multicolumn{1}{c|}{\textbf{Example}}  \\
\hline
 Definition of use cases. & 
 Identify different use cases and applications suitable with QKD which may benefit from its use. & 
 Securisation of data centers supporting virtualised environments, securisation of 5G/6G infrastructures, etc. \\
\hline
Use of standardised interfaces. &
Implement standardised APIs for key exchange, control and management to ensure security, efficiency and compatibility. &
Implementation of ETSI 004, ETSI 014 or ETSI 015 APIs. \\
\hline
Adaptation of standard protocols. &
Adapt the protocols used in the selected application, if needed, in order make use of QKD keys. &
Make changes in protocols such as TLS, IPsec, SSH, etc. to adapt them to the specific features of the QKD keys. \\
\hline
Secure key storage and management. &
Implement proper methods to ensure secure storage and management of QKD keys. &
Use of secure enclaves, strong authentication methods in interfaces, etc.\\
\hline
Secure and efficient use of keys. &
Implement proper methods to ensure that applications make use of keys in a secure and efficient manner. &
Implementation of key-synchronisation methods, definition of re-keying processes, definition of key derivation or re-utilisation politics, etc.\\
\hline
\end{tabular}%
}
\end{table*}

\subsection{Evolution of I2Basque towards a Quantum Internet}
The transition from classical networks to quantum-based communication networks is a challenging task that requires careful analysis of the existing infrastructure. In this case, the current I2Basque network must be studied in detail, with special emphasis on those aspects that may condition the proper integration of quantum technologies. Important aspects requiring careful consideration encompass the number of nodes, link distances, availability of a dark fiber infrastructure, and the multiplexing technology employed to enable the coexistence between classical and quantum signals. 

The conclusions of our initial studies reveal that the topology and technologies employed in the I2Basque network present certain attributes that make it an optimal platform for the transition to a quantum-based infrastructure, as described as follows. Considering the current unavailability of more advanced quantum devices that enable long distance connections, most of the \textbf{link distances} within I2Basque allow the deployment of a quantum network with end-to-end connections. Consequently, the need for intermediate trusted relays, for example, can be avoided. Regarding the transmission of quantum signals, it is worth mentioning the \textbf{absence of a dark fiber infrastructure} in the main ring of I2Basque, so multiplexing techniques must be employed. These techniques enable the transmission of quantum signals alongside the classical ones, without compromising the performance of either communication. In this particular case, as \textbf{DWDM} is the technology employed in the I2Basque network, the feasibility and implications of transmitting quantum signals using this technique must be studied in detail, in order to assess its compatibility, limitations and requirements.

In addition, there are some other attributes that also require further consideration in the design of the new quantum-enabled infrastructure. This includes exploring the need for developing \textbf{new protocols or interfaces} specifically tailored to quantum networks that address their unique requirements. Additionally, \textbf{scalability} is another critical attribute, as it is essential to establish strategies that support possible future growths and evolutions of the network, considering that more advanced quantum devices and capabilities are expected to be integrated, and that the network itself may also be expanded. 

Therefore, the assessment of the main requirements for quantum network implementations according to the State of the Art, show that the I2Basque network has favorable characteristics to support the establishment of a quantum network in its current state and in coexistence with the existing infrastructure. However, more research needs to be done to evaluate all requirements and to design an infrastructure that ensures successful integration of quantum technologies in the I2Basque infrastructure.

\subsection{Deployment of QKD-based secure services}

Considering that QKD has been identified as the key enabling technology for the initial phases of QInternet deployments, the integration of QKD into real-world scenarios and applications stands as an important subject within the SareQuant project. Enabling QKD-based services in current applications as an initial step, in addition to bringing numerous security-related advantages, allows the acquisition of essential knowledge and the laying of the foundations for future quantum services. This strategic approach allows leveraging currently available quantum capabilities, built upon existing systems, while simultaneously paving the way for the adoption of more advanced quantum technologies.

Consequently, several factors have been identified as crucial to be considered and analysed in any integration of QKD into real applications. Table \ref{tab:tabla} presents an overview of these factors, complemented with a short description and some examples for enhanced comprehension. The assessment of all these factors enables a successful integration of QKD into practical applications and the development of robust services.

\section{\uppercase{Conclusions and future work}}
Quantum Technologies are expected to revolutionise current networks and services, with more advanced and promising capabilities. In this context, being NRENs crucial infrastructures for the research and education community, it is essential for them to embrace these technologies in order to enhance their capabilities and contribute to the development of the QInternet. 

This paper presents the SareQuant project, focusing on the design requirements for the evolution of the current I2Basque network into a quantum-based one. The conclusions of the performed analysis up to the date show that the features of I2Basque pose it as an optimal infrastructure to be integrated with quantum technologies. In addition, considering QKD the technology for initial stages of the QInternet, this paper also presents an analysis of requirements and key factors to be considered when integrating QKD into real-world scenarios. This contributes to complete successful and robust integration of QKD with real applications, thus setting the ball rolling towards the adoption of more advanced future quantum technologies. 

Building upon these first steps, there is still more work to be done. Further research is required to explore full potential of quantum networks and its applications, as well as to define all requirements in order to achieve full integration of classical and quantum networks.

\section*{\uppercase{Acknowledgements}}
This work was supported in part by Basque Government through the SareQuant project from IKUR (IKUR-DIPC-PRTR-23/02) and the QFirst project (KK-2022/00062), and in part by the European Commission through the GN5-1 HORIZON-INFRA-2022-NET-01-SGA project.

\bibliographystyle{IEEEtran}

\begin{thebibliography}{1}
\providecommand{\url}[1]{#1}
\csname url@samestyle\endcsname
\providecommand{\newblock}{\relax}
\providecommand{\bibinfo}[2]{#2}
\providecommand{\BIBentrySTDinterwordspacing}{\spaceskip=0pt\relax}
\providecommand{\BIBentryALTinterwordstretchfactor}{4}
\providecommand{\BIBentryALTinterwordspacing}{\spaceskip=\fontdimen2\font plus
\BIBentryALTinterwordstretchfactor\fontdimen3\font minus
  \fontdimen4\font\relax}
\providecommand{\BIBforeignlanguage}[2]{{%
\expandafter\ifx\csname l@#1\endcsname\relax
\typeout{** WARNING: IEEEtran.bst: No hyphenation pattern has been}%
\typeout{** loaded for the language `#1'. Using the pattern for}%
\typeout{** the default language instead.}%
\else
\language=\csname l@#1\endcsname
\fi
#2}}
\providecommand{\BIBdecl}{\relax}
\BIBdecl

\bibitem{van2014quantum}
R.~Van~Meter, \emph{Quantum networking}.\hskip 1em plus 0.5em minus 0.4em\relax
  John Wiley \& Sons, 2014.

\bibitem{rfc9340}
W.~Kozlowski, S.~Wehner, R.~V. Meter, B.~Rijsman, A.~S. Cacciapuoti,
  M.~Caleffi, and S.~Nagayama, ``{Architectural Principles for a Quantum
  Internet},'' RFC 9340, Mar. 2023.

\bibitem{irtf-draft-use-cases}
C.~Wang, A.~Rahman, R.~Li, M.~Aelmans, and K.~Chakraborty, ``{Application
  Scenarios for the Quantum Internet},'' Internet Engineering Task Force,
  Internet-Draft draft-irtf-qirg-quantum-internet-use-cases-16, May 2023, work
  in Progress.

\bibitem{wehner2018QI}
S.~Wehner, D.~Elkouss, and R.~Hanson, ``Quantum internet: A vision for the road
  ahead,'' \emph{Science}, vol. 362, no. 6412, p. eaam9288, 2018.

\bibitem{Survey2018QKD}
Y.~Cao, Y.~Zhao, Q.~Wang, J.~Zhang, S.~X. Ng, and L.~Hanzo, ``The evolution of
  quantum key distribution networks: On the road to the qinternet,'' \emph{IEEE
  Communications Surveys \& Tutorials}, vol.~24, no.~2, pp. 839--894, 2022.

\bibitem{MadQCI}
D.~Lopez, J.~P. Brito, A.~Pastor, V.~Mart{\'\i}n, C.~S{\'a}nchez, D.~Rincon,
  and V.~Lopez, ``Madrid quantum communication infrastructure: a testbed for
  assessing qkd technologies into real production networks,'' in \emph{Optical
  Fiber Communication Conference}.\hskip 1em plus 0.5em minus 0.4em\relax
  Optica Publishing Group, 2021, pp. Th2A--4.

\end{thebibliography}

\IEEEpeerreviewmaketitle

\end{document}